\documentstyle[12pt,epsfig]{article} \textwidth=17cm
\newcommand{\be}{\begin{equation}} \newcommand{\ee}{\end{equation}}
\newcommand{\bea}{\begin{eqnarray}} \newcommand{\eea}{\end{eqnarray}}
\begin{document}

\begin{center}
{\Large \bf Transverse Momentum Dependence of Anomalous\\ {\boldmath
         $J/\psi$} Suppression in  Pb-Pb Collisions}\\[5mm] J\"org
         H\"ufner\\ Institute for Theoretical Physics, University of
         Heidelberg, D-69120 Heidelberg, Germany\\[1mm] and\\[1mm]
         Pengfei Zhuang\\ Physics Department, Tsinghua University,
         Beijing 100084, China \\[5mm]
\end{center}

\begin{abstract}
\setlength{\baselineskip}{16pt}
\noindent The recently published data for $\langle p^2_t\rangle$  for
$J/\psi$ production in Pb-Pb collisions at  158 A GeV are
analyzed. For low values of transverse energy $E_t$, where normal
suppression  dominates,  $\langle p^2_t\rangle (E_t)$ scales with the
path length of  the gluons which fuse to make the $J/\psi$. In the
$E_t$ domain of anomalous suppression $\langle p^2_t\rangle (E_t)$ is
found to rise linearly with the  relative amount of anomalous
suppression. This empirical law is reproduced within an  analytically
solvable transport model which allows high $p_t$ $J/\psi$'s to escape
anomalous  suppression. Interpreted in this way,  the data for
$\langle p^2_t\rangle (E_t)$  lead to an estimate of $t_A \sim 4$
fm/$c$ for the duration of anomalous suppression.

\end{abstract}

\vspace{0.3in}
 
\setlength{\baselineskip}{18pt}

New data mean new surprises. This has been a recurrent phenomenon in
the study of charmonium suppression in high energy nuclear collisions
during the last decade of research.  It also holds for the recently
released data of transverse momentum distributions for charmonia
taken in Pb-Pb collisions at 158 A GeV \cite{na50}. In particular for
central collisions where  anomalously large suppression has been
observed,  the new data display interesting features:   the mean
values  $\langle p^2_t\rangle$ of transverse momentum $p_t$ increase
slowly with transverse energy  $E_t$ and turn steeply upward where the
$J/\psi$ yield drops  (see Fig. 1). Can this behavior tell us something
new about the nature of anomalous $J/\psi$ suppression?

\begin{figure}
\centerline{\includegraphics{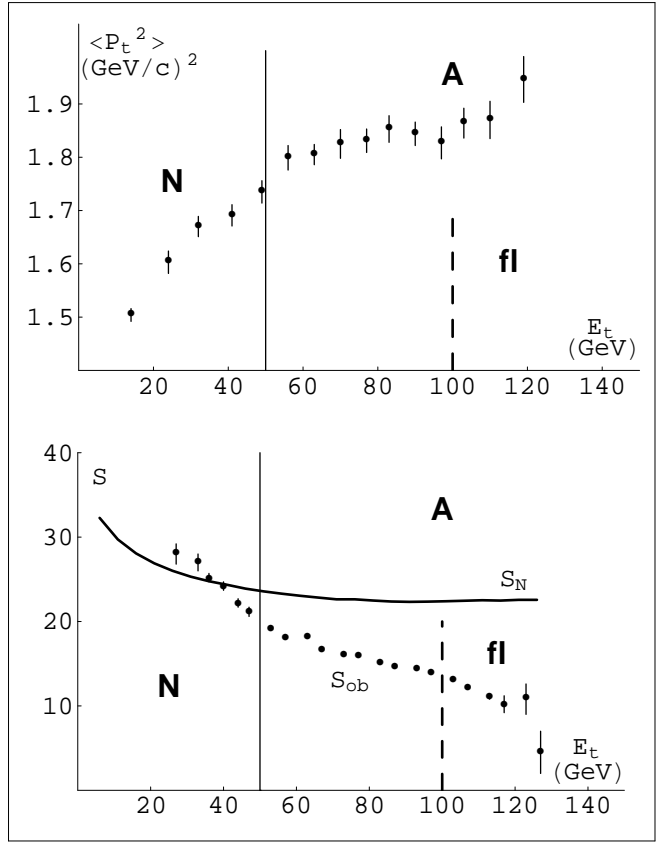}}
\caption{Data for the mean transverse momentum $\langle p_t^2\rangle$
  of $J/\psi$'s produced in $Pb - Pb$ collisions at 158 A GeV as a
  function of transverse energy $E_t$ together with the data for the
  relative $J/\psi$ production cross section $S=\sigma(Pb+Pb\to
  J/\psi+X)/\sigma(Pb+Pb\to DY+X)$ in the same
  experiment\cite{na50}. The range of $E_t$ values is divided into the
  domains of normal suppression (``N'') ($E_t<$ 50 GeV) and anomalous
  suppression (``A'') ($E_t>$ 50 GeV). Within the domain ``A'',  the
  subdomain ``fl'' with $E_t >$ 100 GeV indicates events of high $E_t$
  which rise from fluctuations. The solid line $S_N(E_t)$ is calculated
  for normal nuclear suppression ``N''\cite{blaizot3}, while $S_{ob}$
  denotes the observed values.
}
\end{figure}

The previous data for the  transverse momentum dependence of $J/\psi$
production in  nuclear pA and AB collisions and their interpretation
are presented in two reviews \cite{vogt,gerschel}.  Basically, two
mechanisms are proposed:
\begin{description}
\item[(i)] Rescattering of gluons in the initial state
\cite{gavin1,hufner,blaizot1}: In a pA collision, the gluon  of the
projectile proton scatters from target nucleons before it fuses with a
gluon from the target to form the $J/\psi$. Gluon rescattering in the
initial state  is treated as a random walk in transverse momentum and
the observed   $\langle p^2_t\rangle$ is predicted to increase
linearly with the mean length $\ell_g$ of the path of  the incident
gluon. In an AB collision both gluons which fuse to the $J/\psi$  are
affected by the rescattering. This effect in the initial state has
been  clearly identified in the data, also for AB collisions
\cite{na50}. For central AB  collisions it had been predicted
\cite{gavin2,kharzeev} that  initial state rescattering together  with
anomalously strong $J/\psi$ absorption in the final state may lead to
a saturation or even a decrease of $\ell_g$ and therefore of $\langle
p^2_t\rangle$  for the very large values of $E_t$. The  new data for
Pb-Pb collisions contradict this prediction.
\item[(ii)]     Escape of high $p_t$ $J/\psi$'s in the final state
\cite{matsui,blaizot2,karsch}: Within the scenario of the  quark-gluon
plasma (QGP) anomalous charmonium suppression  occurs in  a limited
space-time region during final state interaction. Only, $J/\psi$'s
with sufficiently high transverse momenta $p_t$ have a chance to
escape the  ``deadly'' region. Therefore anomalous suppression acts
preferentially on  low $p_t$ $J/\psi$'s and the surviving charmonia
should show higher values of  $\langle p^2_t\rangle$ with increasing
amount of  anomalous suppression.  This is what  indeed is seen in the
new data. Do the data then confirm this mechanism?
\end{description}
 
In this letter we analyze the new data for  $\langle p^2_t\rangle$ of
 $J/\psi$ production in Pb-Pb collisions and identify two empirical laws:
 (a) In the domain of normal $J/\psi$ suppression the values for
 $\langle p^2_t\rangle$ depend linearly on the path lengths of the
 gluons.
 (b) A linear correlation between the values of  $\langle
 p^2_t\rangle$ and the relative amount of  anomalous suppression is
 discovered in the $E_t$ range where  anomalous suppression has been
 identified. To our knowledge the second empirical law has not been
 noticed before. We discuss various explanations and then present an
 analytically solvable transport model based on mechanism (ii),
 i.e. escape of high $ p_t$ $J/\psi$'s from the region of anomalous
 suppression. A more detailed analysis of all the data  from
 \cite{na50} in the light of both mechanisms (i) and (ii) is in
 preparation.

Fig. 1 shows the data of ref. \cite{na50}  for  $\langle p^2_t\rangle
(E_t)$ of $J/\psi$'s as a function of transverse  energy $E_t$
together with the previously published data on $J/\psi$ suppression in
the form of the  ratio $S(E_t) =\sigma (Pb+Pb\to J/\psi +X)/\sigma
(Pb+Pb\to DY+X )$. We divide the $E_t$ region into three domains: (a)
Small  transverse energy $(E_t < 50$ GeV), where there is no anomalous
suppression and the yield is  well described by the Glauber approach
with an effective $J/\psi$ nucleon absorption cross section (which
correctly describes pA and S-U collisions).  We denote quantities in
this  domain by the index ``$N$'' (for ``normal''). (b) For values
$E_t > 50$ GeV one observes  anomalous suppression, i.e. the data
$S_{ob}(E_t)$ deviate from the the predictions of the Glauber
approach. We denote all quantities in this $E_t$ region by the index
``$A$'' (for  ``anomalous''). (c) Within the anomalous region the data
at very high values, $E_t > 100$ GeV,  show a particular behavior: the
suppression drops while the  data for  $\langle p^2_t\rangle$
rise. These high values of $E_t$  correspond to the most
central collisions and are interpreted \cite{capella,blaizot3} to
arise from fluctuations  in the transverse energy.  We have indicated
this origin by the symbol ``$fl$'' at the appropriate  places.

We begin our analysis of the data by investigating to which degree
(for which values  of $E_t$) the mechanism (i), gluon rescattering in
the initial state, explains  the  data. This
mechanism leads to a dependence \be\label{pt2ab} \langle
p^2_t\rangle^{AB}(E_t)= \langle p^2_t\rangle^{NN}+
\frac{\langle p^2_t\rangle^{gN}}{\lambda_{gN}}\ell_g^{AB}(E_t)\ , \ee
where  $\langle p^2_t\rangle^{NN}$ is the contribution already present
in $J/\psi$ production in the elementary  $NN$ event, while the second
term is linear in the mean length $\ell_g(E_t)$, which the two gluons
travel in nuclear matter before they fuse. The constant in front of
$\ell_g$ depends on $\langle p^2_t\rangle^{gN}$, the mean transverse
momentum  acquired in a gluon-nucleon collision, and $\lambda_{gN}$,
the mean free path of a gluon in  nuclear matter. This constant is
taken as an adjustable parameter, while $\ell_g(E_t)$ is
calculated. Then the observed values of $\langle
p^2_t\rangle^{AB}(E_t)$ are plotted  versus the calculated values of
$\ell_g^{AB}(E_t)$. If a linear relation appears, one takes it as
support  for the gluon rescattering mechanism. This analysis has
already been done in ref. \cite{na50}  for the  Pb-Pb data. However,
$\ell_g^{AB}(E_t)$ is calculated neglecting absorption of the $J/\psi$
in the final  state. We therefore repeat their analysis with absorption.
 Of course, the results   depend on the employed
absorption model. We choose the one of ref. \cite{blaizot3}, which
correctly  reproduces the data for $J/\psi$ suppression even in the
domain of fluctuations. Then \be\label{algab}
\ell_g^{AB}(E_t)=\frac{\int d^2 b\  d^2s\ dz_A\
dz_B\left(\ell_g^A(\vec s,z_A)+\ell_g^B(\vec b -\vec
s,z_B)\right)K(\vec b,\vec s,z_A,z_B,E_t)} {\int d^2b\ d^2s\ dz_A\
dz_B\ K(\vec b,\vec s,z_A,z_B,E_t)}\ , \ee where \bea\label{lgab} &&
\ell^A_g(\vec s,z_A)=\int^{z_A}_{-\infty}dz\rho_A(\vec s,z)/\rho_0\ ,
\nonumber\\ && \ell^B_g(\vec b -\vec
s,z_B)=\int^{\infty}_{z_B}dz\rho_B(\vec b -\vec s,z)/\rho_0\ \eea are
the geometric lengths which the two gluons travel through the nuclear
density along the $z$-direction   with impact parameter $b$, and
$\rho(\vec r)$ is the density of nuclear  matter. To simplify the
numerical calculations, we use in the following the uniform
distribution.  The expression for the kernel $K$ in eq. (\ref{algab})
is taken from ref. \cite{blaizot3}:  \bea\label{kernel} K(\vec b,\vec
s,z_A,z_B,E_t)&=&\rho_A(\vec s,z_A)\rho_B (\vec b -\vec
s,z_B)\cdot\nonumber\\ &&\exp\left(-\sigma^{J/\psi
N}_{abs}\left[\int^\infty_{z_A}dz\rho_A(\vec
s,z)+\int_{-\infty}^{z_B}dz\rho_B(\vec b -\vec s,z)\right]\right)\cdot
\nonumber\\ &&\Theta\left(n_c-\frac{E_t}{\langle E_t\rangle(\vec
b)}n_p(\vec b,\vec s)\right)\cdot P(E_t|\vec b).\eea Here, the final
state absorption for the $J/\psi$ is contained in the exponential
(``normal''  absorption by  nucleons with an absorption cross section
$\sigma_{abs}^{J/\psi}$) and in the Theta function  (``anomalous''
suppression suddenly setting in, when the density of produced matter
exceeds a  threshold density $n_c$). The function $P(E_t|b)$ describes
the distribution of transverse energy in  events at a given impact
parameter $b$. We have followed ref. \cite{blaizot3} in the notation
and the  numerical values for the constants and therefore refer the
reader who is interested in more details,  to this paper.

\begin{figure}
\centerline{\includegraphics{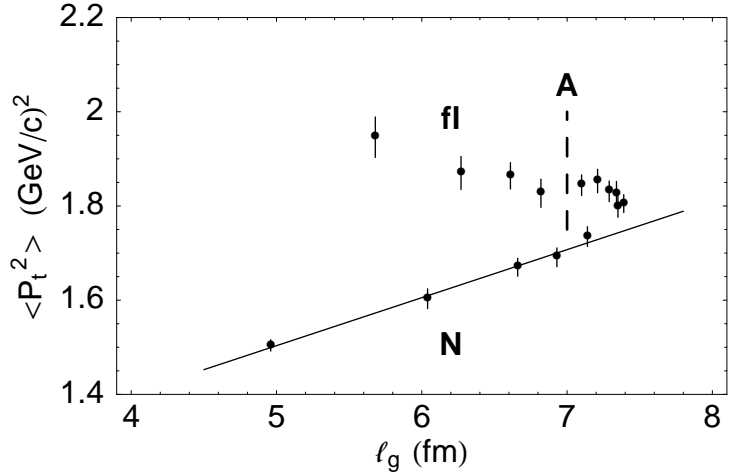}}
\caption{The observed values $\langle p_t^2\rangle(E_t)$ for
  $J/\psi$'s from $Pb - Pb$ collisions \cite{na50}
  plotted versus $\ell_g(E_t)$, the
  mean path of the gluons before they fuse to the $J/\psi$. The values
  $\ell_g(E_t)$ are calculated within the model\cite{blaizot3} which
  contains normal and anomalous nuclear suppression and which is able to
  reproduce the cross section data in Fig.1. As defined in Fig.1 we have
  divided the $E_t$ range into domains ``N'', ``A'' and ``fl'': Only the
  five data points on the straight line belong to the domain of normal
  suppression, while all data on the backward branch belong to anomalous
  suppression. 
}
\end{figure}

Fig.2 shows a plot of the experimental values of  $\langle
p^2_t\rangle(E_t)$ versus the calculated  values of
$\ell_g^{AB}(E_t)$. The qualitative picture is rather different from
the one given in ref. \cite{na50}: Only  those values of $E_t$, which
correspond to the domain of normal (``$N$'') suppression in Fig. 1,
show a  linear relation.
We fit it by a straight line and find a
  value for the slope constant
\be\label{agn} a_{gN} = \frac{ \langle
p^2_t\rangle_{gN}}{\lambda_{gN}} =\left(0.102\pm 0.006\right)
(GeV/c)^2/fm,\ee which differs from the value $ (0.081 \pm0.003)$
(GeV/$c)^2$/fm given in ref. \cite{na50}, since we use a 
 different prescription for the
calculation of  $\ell_g^{AB}$. In the $E_t$ domain  where
anomalous suppression (``$A$'') sets in, the plot shows an
anticorrelation: With increasing values of $E_t$, the calculated
values of $\ell_g(E_t)$ decrease, while the experimental values for
$\langle p_t^2\rangle (E_t)$ increase.  This behavior in the anomalous  domain
hides some physics other than gluon rescattering and will be studied in
the following.

 We assume that all $J/\psi$'s which are suppressed by an anomalous
mechanism  have  still another source influencing the
transverse momentum distribution above the one from gluon rescatering
in the initial state. If we decompose the observed cross section
$S_{ob}(E_t)$  for $J/\psi$-production 
(relative to $DY$-production)
into its two contributions
\be\label{sob} S_{ob}(E_t)=S_N(E_t)-S_A(E_t),\ee where the
contribution $S_N(E_t)$ from normal absorption is a theoretical
quantity and is  calculated by using the kernel $K$ from
eq. (\ref{kernel}), {\it without} the Theta function and is shown in  Fig. 1
by the solid line.  The anomalous suppression, $S_A(E_t)$, is defined
as the difference between observed,  $S_{ob}(E_t)$, and calculated
values, $S_N(E_t)$.  We associate different values,
$\langle p^2_t\rangle_N(E_t)$ and  $\langle p^2_t\rangle_A(E_t)$, 
 with the normal and anomalous  contributions, respectively.
 Then the observed value  $\langle
p^2_t\rangle_{ob}(E_t)$  can be written as
 \bea\label{pt2ob} 
\langle p^2_t\rangle_{ob}(E_t)&=&\frac{S_N(E_t) \langle
p^2_t\rangle_N(E_t)-S_A(E_t) \langle p^2_t\rangle_A(E_t)}
{S_{ob}(E_t)}\nonumber\\ &=&  \langle
p^2_t\rangle_N(E_t)+\frac{S_A(E_t)}{S_{ob}(E_t)} \delta
p^2_t(E_t),\eea
 where
 \be\label{dpt2} 
\delta p^2_t(E_t)= \langle
p^2_t\rangle_N(E_t)- \langle p^2_t\rangle_A (E_t).\ee
Eq. (\ref{pt2ob}) should be valid in the full $E_t$ domain comprising
anomalous suppression and 
normal one (where $S_A(E_t)=0$, by definition).
  For  normal suppression, the $E_t$ dependence of the observed
values $\langle p_t^2\rangle_{ob}(E_t)$ is solely carried by values
$\langle p^2_t\rangle_N(E_t)$ calculated from eqs. 
(\ref{algab}-\ref{kernel}) without the $\Theta$-function and displayed
in Fig. 2. The calculation of $\langle p^2_t\rangle_N(E_t)$ can also
be extended into the domain of anomalous suppression $(E_t>50$
GeV). We find  that $\langle p^2_t\rangle_N(E_t)$ changes by maximally
3\% over the whole domain $0.25\leq S_A/S_{ob}\leq 1.25$, where data
exist and are displayed in Fig. 3. Therefore, in the domain of
anomalous suppression, the $E_t$ dependence of $\langle
p_t^2\rangle_{ob}(E_t)$ is predominantly carried by the second term in
eq. (\ref{pt2ob}). In order to disentangle the $E_t$ dependence
residing in $S_A/S_{ob}$ from the one in $\delta p^2_t(E_t)$, we plot
 the observed values of
$\langle p^2_t\rangle_{ob}(E_t)$  versus the ratio
$S_A(E_t)/S_{ob}(E_t)$. The result  is displayed in Fig. 3. To a very
good accuracy all points lie on a straight line with a slope
\be\label{dpt22} \delta p^2_t=(0.132\pm 0.007) (\rm{GeV}/c)^2.\ee
The result indicates that the $E_t$ dependence of  
$\langle p^2_t\rangle_{ob}(E_t)$ in the domain of anomalous
suppression arises predominantly from the ratio $S_A(E_t)/S_{ob}(E_t)$.
 We also draw the attention to the $E_t$ region ``$fl$'', where
fluctuations dominate: While a sudden  drop in
 the $J/\psi$ suppression and a sudden rise in  $\langle
p^2_t\rangle$ occur (Fig. 1),  the empirical regularity shown
in Fig. 3 continues into the domain ``$fl$'' without any noticeable
change in character indicating that no new physics appears in this
domain of $E_t$.
 The  empirical correlation displayed in Fig. 3
is the first new result of this paper.

\begin{figure}
\centerline{\includegraphics{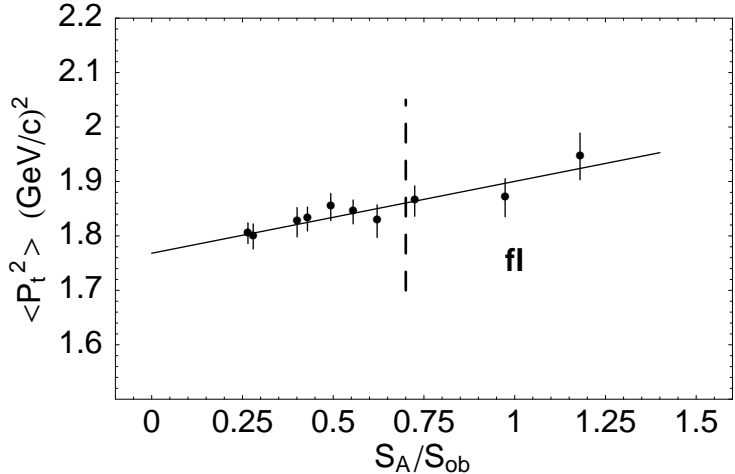}}
\caption{The observed values $\langle p_t^2\rangle(E_t)$ (in the
  ``A'' domain, $E_t>$ 50 GeV) for $J/\psi$'s from $Pb-Pb$ collisions
  plotted against the relative amount of anomalous suppression
  $S_A(E_t)/S_{ob}(E_t)$. Here, $S_A(E_t)$ is the difference between the
  calculated value $S_N(E_t)$ and the observed values $S_{ob}(E_t)$ in
  Fig.1. (The data in the ``N'' domain, not shown, would all be found at
  $S_A/S_{ob}=0$).
}
\end{figure}

Can we understand this relation?  Are we able to calculate the
constant $\delta p^2_t$? We will discuss two possible explanations,
intermediate $\chi$ production and  escape of high $p_t$ charmonia in
the final state and start with the $\chi$.

 The observed intensity of $J/\psi$'s produced in nuclear collisions
 has two contributions,  directly produced $J/\psi$'s and those
 arising from produced $\psi'$'s and $\chi$'s  which decay  into
 $J/\psi$ long after the collision but before detection.  The indirect
 contributions amount to  about 40\% \cite{vogt,gerschel}. Since
 $\psi'$'s and $\chi$'s are less bound than the $J/\psi$, it has been
 argued \cite{matsui} that anomalous  suppression should act
 predominantly on the $\chi$'s and $\psi'$'s rather  than on the
 directly produced $J/\psi$'s. The $\chi$'s may have a smaller value
 of   $\langle p^2_t\rangle_\chi$ than  the $J/\psi$ and $\psi'$,
 since the $\chi$'s can be produced directly by fusion of two gluons.
  $J/\psi$ and $\psi'$ production involves three gluons, where  the
 third one is presumably radiated off  from a color octet intermediate
 state and generates additional $\langle p^2_t\rangle$.  If anomalous
 suppression acts predominantly on the $\chi$,  the linear relation
 eq. (\ref{pt2ob}) seems plausible  with 
\be\label{dpt23} \delta
 p^2_t\sim \left(\langle p^2_t\rangle_{J/\psi}^{NN}- \langle
 p^2_t\rangle^{NN}_\chi\right).\ee
Unfortunately, we have no 
calculations for the difference in eq. (\ref{dpt23}).
  Furthermore, the emission of a gamma $(\chi
 \to J/\psi + \gamma)$ reduces the difference by about $0.1 (GeV/c)^2$
 and may even change the sign of $\delta p_t^2$. If  the mechanism of
 intermediate $\chi$ production was the dominant explanation for
 the observed value of  $\delta p^2_t$, the values of  $\langle
 p^2_t\rangle$   observed in the production of $\psi'$'s should have
 no anomalous values.  However, the  contrary is true according to the
 data in Pb-Pb collisions \cite{na50}. Therefore the  contribution of
 $\chi$ to the observed $J/\psi$ does not seem a compelling
 explanation for the  behavour of  $\langle p^2_t\rangle$ in the
 anomalous region.

We come to the second explanation: Only  high $p_t$ charmonia
 escape anomalous suppression.
This argument  is
 not new\cite{matsui,blaizot2,karsch}. In this paper we start from
 this idea and propose a
 solvable transport model, in order to see whether
 the empirical relation Fig. 3 and the value $\delta p_t^2$,
 eq. (\ref{dpt22}), can be understood.

We introduce the phase space distribution $n(\vec r,\vec p,t)$ for
$J/\psi$'s produced in a Pb-Pb  collision with a definite value of
$E_t$.  In a system where the $J/\psi$'s have zero longitudinal
momentum, the values of  $\vec r$ and $\vec p$ denote the transverse
position and momentum of the $J/\psi$'s,  respectively. We denote by
$t=0$ the time, when all  collisions involving nucleons (production
$N+N\to J/\psi + X$ and suppression $J/\psi +N\to$ no $ J/\psi$) have
ceased. Then at $t=0$ we have what we call   normal suppression and
therefore have \be\label{sn} S_N=\int d^2\vec r\ d^2\vec p\ n(\vec
r,\vec p,0)\ ,\ee \be\label{pt2n} \langle p^2_t\rangle_N=\int d^2 \vec
r\ d^2\vec p\ p^2\ n(\vec r,\vec p,0)/S_N\ee
 for the intensity $S_N$
and  $\langle p^2_t\rangle_N$ of the $J/\psi$, respectively.
Here and in the following we consider one particular impact parameter
(one value of $E_t$).
 For $t>0$
the phase space  distribution $n(\vec r,\vec p,t)$ evolves under the
influence of  anomalous suppression. In order to obtain analytical
results, we have to introduce three simplifying assumptions:  (a) Instead of
the absorption process happening continuously for $t>0$ until all
charmonia are ``eaten up'', we let anomalous suppression
 happen at one particular time
$t=t_A$, while for $0<t<t_A$, the phase space  distribution  evolves
freely, i.e.  \be\label{nrpt} n(\vec r,\vec p,t)=n\left((\vec r+\vec v
t),\vec p,0\right),\ee where $\vec v = \vec p/M$ is the velocity of a
$J/\psi$ with momentum  $\vec p$ within a non-relativistic
approximation. (b) Anomalous suppression happens in such a way that
all $J/\psi$'s,  whose position $r$ is smaller than  a given radius
$r_A$ are suppressed (where $r_A$ depends on $E_t$,  $r_A(E_t)$). This
assumption corresponds to the  $\Theta$ function, eq. (\ref{kernel}),
introduced in ref. \cite{blaizot3}.  Then    \be\label{sob2}
S_{ob}(E_t)=\int_{r<r_A(E_t)}d^2\vec r\ d^2\vec p\ n(\vec r,\vec
p,t_A)\ee 
\be\label{pt2ob2} \langle
p^2_t\rangle_{ob}(E_t)=\int_{r<r_A(E_t)}d^2\vec r\ d^2\vec p\ p^2\ n(\vec
r,\vec p,t_A)/S_{ob}(E_t)\ee for the observed intensity $S_{ob}(E_t)$
and the observed  $\langle p^2_t\rangle_{ob}(E_t)$, respectively. With
the  further assumption (c) of a Gaussian for the initial phase space
distribution \be\label{nrpt2} n(\vec r,\vec
p,t=0)=c_0\exp\left(-\frac{r^2}{R^2}-\frac{p^2} { \langle
p^2_t\rangle_N}\right)\ee all integrals in
eqs. (\ref{sn}-\ref{pt2ob2}) can be evaluated explicitly and after
some algebra one  arrives at \be\label{pt2ob3} \langle
p^2_t\rangle_{ob}(E_t)= \langle p^2_t\rangle_N+\ln
\left(1+\frac{S_A(E_t)}{S_{ob}(E_t)}\right)\cdot\delta p^2_t\ ,\ee
\be\label{dpt24} \delta p^2_t= \langle p^2_t\rangle_N\cdot
\frac{\langle p^2_t\rangle_N}{M^2} t^2_A R^{-2}.\ee

Eq. (\ref{pt2ob3}) for the mean squared transverse momentum $\langle
  p^2_t\rangle_{ob}(E_t)$ in the domain of anomalous suppression  is
  the second new result of our paper. It shows the dependence of this
  quantity on the degree  of anomalous suppression $S_A = S_N-S_{ob}$,
  and one recovers the empirical law, eq. (\ref{pt2ob})  after
  expanding the logarithm for small values of $S_A/S_{ob}$ (the
  logarithmic  dependence seems to be an artifact of the Gaussian
  shape for the phase space distribution eq. (\ref{nrpt2})).
  Furthermore, eq. (\ref{dpt24}) gives an analytical expression for
  the value  $\delta p_t^2$.  It involves the mean squared  transverse
  velocity of the produced $J/\psi$, the mean squared transverse
  radius $R$ of the overlap  zone of the two colliding nuclei (for
  central collisions, $R$ is related to the mean squared radius of the
  Pb nucleus) and the time $t_A$, when anomalous suppression happens.
  Using the empirical values for $\delta p^2_t = 0.132 (GeV/c)^2$,
  eq. (\ref{dpt22}),  $\langle p^2_t\rangle_N= 1.77 (GeV/c)^2$  and
  $R^2={2\over 5}\langle r^2\rangle_{Pb}$, one finds $t_A= 4$
  fm/$c$.  This value is not unreasonable in view of  values of $5-7$
  fm/$c$  discussed in the literature\cite{rapp} for the lifetime of
  the  fireball in central Pb-Pb collisions. We expect  our value for
  $t_A$ to change somewhat, if some of the above assumptions (a-c) are
  relaxed.  However, the basic dependence of $\delta p^2_t$ on the
  quantities $R$,  $\langle p^2_t\rangle_N$ and $t_A$ should  remain
  (also for dimensional reasons).

In summary, we have been able to show, that {\it normal}
 nuclear suppression
 goes along  with values of  $\langle p^2_t\rangle$ which depend
 linearly on the path lengths $\ell_g^{AB}$
 of the gluons which fuse to  form the
 $J/\psi$, Fig. 2. In the domain of {\it anomalous} suppression, the data
 show a linear dependence on  the relative amount of anomalous
 suppression $S_A/S_{ob}$. This relation also holds for the domain  of
 fluctuations in $E_t$ and supports the interpretation that no
 qualitatively new mechanism  begins for these very high values of
 $E_t$. The empirical relation in the domain of anomalous suppression
 is consistent within the geometric picture ``high $p_t\: J/\psi$'s
 escape anomalous suppression'', which has been proposed \cite{matsui,
 blaizot2} and worked out \cite{karsch} long ago.

\vspace{0.3in}

\noindent {\bf \underline{Acknowledgments:}} We thank Boris
Kopeliovich for some stimulating discussions. One of  the authors
(P.Z.) is grateful for the hospitality at the Institute for
Theoretical Physics in  Heidelberg, his work was supported by the
grants 06HD954 and NSFC19925519.

\end{document}